\documentclass[preprint,showpacs,preprintnumbers,amsmath,amssymb]{revtex4}

\usepackage{graphicx}
\usepackage{dcolumn}
\usepackage{bm}

\usepackage{epstopdf} 
\newcommand{\g}{\text{g}}
\newcommand{\CN}{\text{CN}}
\newcommand{\F}{\text{F}}
\newcommand{\B}{\text{B}}

\begin{document}

\title{Quantum oscillations observed in graphene at microwave frequencies}

\author{P. Jiang}
\altaffiliation{Current address: Department of Physics, National Taiwan Normal University, Taipei 116, Taiwan}
\affiliation{National High Magnetic Field Laboratory, Tallahassee, Florida 32310, USA}
\affiliation{Princeton University, Princeton, New Jersey 08544, USA}
\author{A. F. Young}
\affiliation{Department of Physics, Columbia University, New York, New York 10027, USA}
\author{W. Chang}
\affiliation{Department of Physics, Columbia University, New York, New York 10027, USA}
\author{P. Kim}
\affiliation{Department of Physics, Columbia University, New York, New York 10027, USA}
\author{L. W. Engel}
\affiliation{National High Magnetic Field Laboratory, Tallahassee, Florida 32310, USA}
\author{D. C. Tsui}
\affiliation{Princeton University, Princeton, New Jersey 08544, USA}


\begin{abstract}
We have measured the   microwave conductance of mechanically exfoliated graphene   at  frequencies up to 8.5 GHz.   The conductance at 4.2 K exhibits quantum oscillations, and is independent of the frequency.  
 \end{abstract}

\maketitle

 The discovery  \cite{Novoselov04} that graphene, a   single atomic layer of carbon,  can be produced by mechanical exfoliation of graphite  has   resulted in an explosion of  research activity \cite{Geim07, Geim09}.  Interest in graphene arises from its  ``Dirac" spectrum, which resembles that of relativistic particles \cite{Novoselov05, Zhang05}.  This spectrum gives rise to characteristic quantum oscillations in the magnetoresistance, and for sufficiently high magnetic fields and low disorder  to the quantum Hall effect \cite{Novoselov04, Novoselov05, Zhang05}. There is also great interest in graphene for potential applications, including microwave   transistors, which were demonstrated \cite{Lin09, Lin10} to have cutoff frequencies up to 100 GHz.   A recent study \cite{delig} of  a graphene-loaded coplanar waveguide at  room temperature and zero   magnetic field showed that   reasonable impedance matching to 50-$\Omega$ systems is possible.

In this paper we present a study of the microwave  two-terminal conductance of mechanically exfoliated graphene. Our measurements are carried out on relatively low-mobility ($\mu\sim 1000$ cm$^2$/V-s) samples  at 4.2 K and  in magnetic fields ($B$) up to 8 T.  Quantum oscillations of the conductance are observed up to our maximum measuring frequency ($f$) of 8.5 GHz, both as a function of magnetic field and carrier density.    To within the acccuracy of our measurement, the conductance remains frequency independent from dc up to 8.5 GHz,  consistent with $2\pi f \tau\ll 1 $,  where $\tau$ is the transport    relaxation time  of   the graphene ($\sim 10$ fsec for our samples).

\begin{figure}
\includegraphics[width=3.25in]{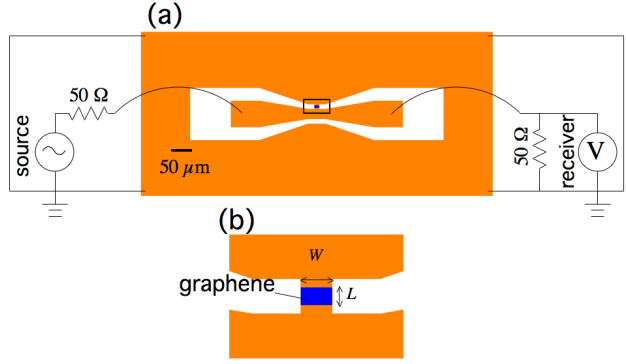}
\caption{\label{fig:fig1}
(a) Schematic of the microwave circuit connected to the graphene device. Orange regions represent the metal film contact pattern deposited on SiO$_{2}$/Si. (b) Expanded view of the graphene flake and the nearby contact pattern connected to it (indicated with a rectangular box in 1a). }  
\end{figure}

Fig.~\ref{fig:fig1} shows a schematic of the measurement set-up.   A graphene flake mechanically exfoliated  from bulk Kish graphite is placed  on top of a SiO$_{2}$/Si substrate, and a Ti:Au metal film is patterned as shown using electron-beam lithography. The graphene flake connects a driven center conductor and a grounded outer  conductor.   The Si substrate is mounted in a fixture which uses $\sim1$ cm long  planar transmission lines between  the substrate and coaxial connectors.  The planar lines in the fixture  are  gold-wire bonded to the metal on the substrate.  The connectors in the fixture    attach to coaxial cables connected  to a room-temperature network analyzer. The graphene conductance ($G$) affects the microwave transmission coefficient, acting essentially as a load across a 50-$\Omega$ transmission line, so that larger $G$ produces lower transmission.  In contrast to most dc experiments \cite{Zhang05, gaowkloc, zhangnu014, lauaspectrat, JHChen08},    the graphene in this set-up is connected at two terminals, so the measured conductance can include effects of contact  resistances.

In order to minimize capacitive losses, we carefully chose the 
room-temperature resistivity of the Si substrate used as gate electrodes in this experiment to be 0.01 to 0.02 Ohm-cm. At the experimental temperature of 4.2 K, the conductivity of this substrate is sufficiently low that the current through the graphene smaple dominates. We verify this condition by measuring   a piece of the same SiO$_{2}$/Si  wafer  with the metal film pattern but without  graphene.  The substrate conductivity is still large enough to allow control of the graphene carrier density and sign  by applying a backgate voltage ($V_\g$) between the graphene and  the substrate.

\begin{figure}
\includegraphics[width=3.25in]{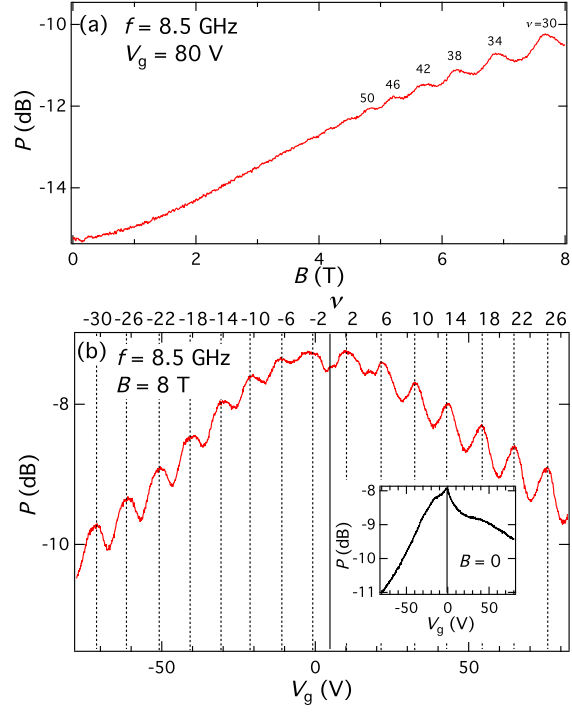}
\caption{\label{fig:fig2}
(a) Log-scaled transmitted power ($P$) as a function of the magnetic field ($B$) at $f=8.5$ GHz and a backgate voltage ($V_\g$) of 80 V, from sample 1.  (b) $P$ vs  $V_\g$ at $f=8.5$ GHz and $B=8$ T, from sample 1. Integer filling factors ($\nu$)  label   dotted vertical lines at the local maxima. The inset shows $P$ vs $V_\g$ at $B=0$ from sample 2. }
\end{figure}

We show results  from  two graphene flakes: sample 1, with  length $L\approx2$ $\mu$m and width $W\approx 42$ $\mu$m,  and sample 2, with $L\approx 1.6$ $\mu$m and $W\approx 15$ $\mu$m.  
   From the measured resistivity at large carrier density,  we found    dc electron and hole mobilities of $0.5$ to $1\times 10^{3}$ cm$^{2}$/V-s, with the hole mobility higher than the electron mobility as has been observed elsewhere \cite{Huard08,blakecontact}. The onset fields of the Shubnikov-de Haas oscillations result in an estimate of mobilities \cite{Bolotin08} $\mu\sim2.2$ to $4\times 10^{3}$ cm$^{2}$/V-s. The mobilities are rather  low  compared to those of typical mechanically exfoliated  graphene samples \cite{Tan07, JHChen08}. This low mobility might be due to contamination during the device fabrication process of the large electrodes for microwave contacts or to the relatively large sample sizes, which tend  to have large  inhomogeneity.   The samples for which we show data have  backgate voltages at the charge neutral point ($V_\CN$) within 5 V of zero.

Fig.~\ref{fig:fig2}a shows $P$, the microwave power transmitted through graphene sample 1, as a function of $B$  at $f=8.5$ GHz  with $V_\g=80$ V so that the graphene electron density is $\sim5.4\times 10^{12}$ cm$^{-2}$.  Quantum oscillations are evident for $B$ above about 4 T. Local maxima in transmission correspond to  local minima  of  conductance, and are taken as integer Landau fillings ($\nu$) as considered in detail in Ref.~\onlinecite{Abanin08}. Fig.~\ref{fig:fig2}b   shows $P$ vs $V_\g$ in a magnetic field of 8 T.   The quantum oscillations are clearly visible in this trace as well, and are superimposed on a broad peak around $V_\CN$.   At $B=0$, the  peak in $P$ vs $V_\g$ around $V_\CN$ is present without the quantum oscillations, as shown in the inset.  

For graphene  admittance ($Y$) sufficiently small in magnitude, the  microwave attenuation of the  fixture   with graphene can be linearized as $ \log |s_{21}/s^0_{21}|= C_F(f)  G $, where $s_{21}$ and $s^0_{21}$   are  transmission coefficient $s$ parameters \cite{Adam69} between the fixture connectors with and without graphene.   $G=\mbox{Re}(Y)$, and $C_\F(f)$ is a frequency-dependent  constant associated with the fixture.  $C_\F$ is obtained from room-temperature measurements of the fixture with standard resistors (carbon paint) painted onto the microwave pattern  shown in   Fig.~\ref{fig:fig1} with an undoped substrate.    Planar simulation software \cite{sonnet} was used to verify that there was no effect of the  geometry of the standard resistor   differing from that of the graphene flake.  From  an analytical  model of the fixture, which took into account the lengths within the fixture, the conductor losses, and the reactances of the bond wires \cite{Sutono01}, we find  the linearized attenuation to hold to within about 5 \%\  for $G\le 9 $ mS  for the case of $Y$  having small argument.  We verify that  $Y$ has small argument for our experimental conditions by  looking at   the phase shift of the microwave transmission  as $V_\g$ is varied.  At any frequency in the experiment this phase shift is within $7^\circ$, which is  within experimental error of that expected for $Y$ purely real. At fixed $f$ we obtain  $\Delta  G =G(V_\g)-G(V_\CN)= C_\F ^{-1} \log (P (V_\g)/P(V_\CN))$.

\begin{figure}
\includegraphics[width=3.5in]{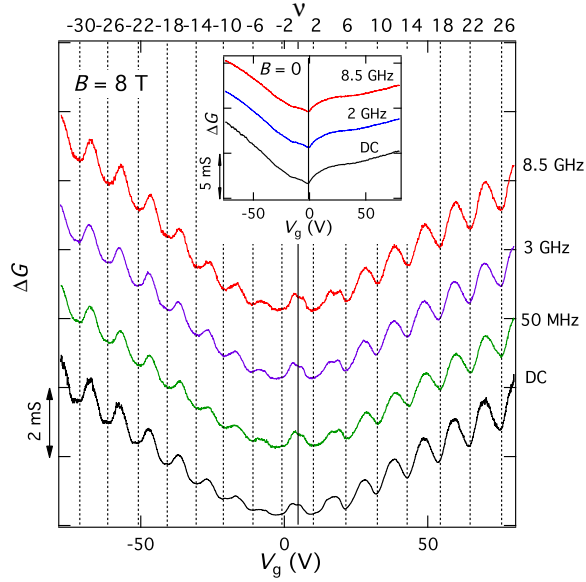}
\caption{\label{fig:fig3}
$\Delta G$, the graphene conductance shift from its charge neutral point value, vs backgate voltage ($V_\g$)  for sample 1 at $B=8$ T,  at various frequencies ($f$) from DC up to 8.5 GHz. Traces are offset for clarity. The charge neutral point voltage, $V_\CN=4.7$ V,  is indicated with a vertical solid line.  The minima of the small quantum oscillations are indicated with vertical dashed lines with their corresponding $\nu$ shown on the top axis. Inset: $\Delta G$ vs $V_\g$ at $B=0$ as measured in  sample 2.}
\end{figure}

Fig.~\ref{fig:fig3} shows $\Delta G$ vs $V_{\g}$  at various frequencies from dc to 8.5 GHz.   The two-terminal dc conductance is measured between the contact center line and the ground plane with a lock-in technique.   The  $B=0$  data shown in the inset is   obtained   from sample 2, and the charge neutral point dc conductance is  $G(V_\CN)=  1.6$ mS.  The  8-T data in the figure is from sample 1, and has dc $G(V_\CN)=0.6 $ mS.  Clear quantum oscillations are apparent, but possibly due to the low mobility of our samples,   $G$ does not appear to exhibit quantized values\cite{Abanin08} as would be expected if the  quantum Hall effect were developed.

  For both data sets in   Fig.~\ref{fig:fig3},  $\Delta G$ vs $V_\g$  is  clearly independent of $f$ within experimental error. At $B=8$ T the maximum percent difference between  microwave  measured $\Delta G(V_\g)$ and the conventionally measured  dc $\Delta G(V_\g)$ is about 20 \% for $V_\g<0$, and 5 \% for $V_\g>0$.  For $B=0$, the  microwave $\Delta G(V_\g)$ is within about 15\% of the dc $\Delta G$ for both positive and negative   $V_\g$. Additional data for lower nonzero $B$ likewise shows no $f$ dependence of $\Delta G$ to within our experimental accuracy.  Though we measure $\Delta G$ only, its   frequency independence is a reliable indicator  that the absolute conductance $G$  must be frequency independent as well.

The $f$-independent behavior of graphene conductance is entirely consistent with the low-$f$ limit, $2\pi f\tau<<1$, where $\tau$ is the   transport  relaxation time   of the graphene carriers.  This is reasonable in light of   estimates of such time scales from theory\cite{Hwang08}  and from dc mobility\cite{Tan07} and magneto transport measurements\cite{Hong09}.  All give  transport and quantum scattering times for mechanically exfoliated graphene of  order 10 fsec for $ n_{c}\ \sim 10^{12}$ cm$^{-2}$ and $\mu\sim 1000$ cm$^2$/V-s. Our microwave frequencies are also too small to resolve  inter Landau level transitions\cite{McClure56, Haldane88, Iyengar07, Jiang07} in the magnetic field.   Our measuring temperature is much larger than $hf/k_\B$ (where $h$ is the Planck's constant and $k_\B$ is  the Boltzmann constant), so that   quantum critical effects \cite{Giesbers09}    are also not expected to result in the measured $f$ dependence  \cite{sondhicarinirmp}.

In summary, we have measured the conductance of  low mobility,  mechanically-exfoliated graphene  devices with contact patterns designed for microwave measurements on the SiO$_2$/Si chip, at frequencies  up to 8.5 GHz and at a temperature of 4.2 K.     The conductance, which at high $B$ includes quantum oscillations vs magnetic field and backgate voltage, is $f$ independent to within the accuracy of our measurements.  We thus conclude that the graphene magnetotransport  remains essentially in the dc limit  up to  at least 8.5 GHz.

 The work at NHMFL was supported by DOE Grant Nos. DE-FG21-98-ER45683. NHMFL is supported by NSF Cooperative Agreement No. DMR-0084173, the State of Florida and the DOE. P. Kim and A. F. Young acknowledge the support from DARPA CERA program and AFOSR MURI.


\begin{thebibliography}{999}

\bibitem{Novoselov04} K. S. Novoselov, A. K. Geim, S. V. Morozov, D. Jiang, Y. Zhang, S. V. Dubonos, I. V. Grigorieva, and A. A. Firsov, Science {\bf 306}, 666 (2004).

\bibitem{Geim07} A. K. Geim and K. S. Novoselov, Nature Mater. {\bf 6}, 183 (2007).
\bibitem{Geim09} A. K. Geim, Science {\bf 324}, 1530 (2009).

\bibitem{Novoselov05} K. S. Novoselov, A. K. Geim, S. V. Morozov, D. Jiang, M. I. Katsnelson, I. V. Grigorieva, S. V. Dubonos, and A. A. Firsov, Nature (London) {\bf 438}, 197 (2005).

\bibitem{Zhang05} Y. Zhang, Y. W. Tan, H. L. Stormer, and P. Kim, Nature (London) {\bf 438}, 201 (2005).

\bibitem{Lin09} Y.-M. Lin, K. A. Jenkins, A. Valdes-Garcia, J. P. Small, D. B. Farmer, and P. Avouris, Nano Lett. {\bf 9}, 422 (2009).
\bibitem{Lin10} Y.-M. Lin, C. Dimitrakopoulos, K. A. Jenkins, D. B. Farmer, H.-Y. Chiu, A. Grill, and P. Avouris, Science, {\bf 327}, 662 (2010).

\bibitem{delig} G. Deligeorgis, M. Dragoman, D. Neculoiu, D. Dragoman, G. Konstantinidis, A. Cismaru, and R. Plana, Appl. Phys. Lett., {\bf 95}, 073107 (2009).

\bibitem{gaowkloc} X. P. Gao, A. P. Mills, Jr., A. P. Ramirez, L. N. Pfeiffer, and K. W. West, Phys. Rev. Lett. {\bf 89}, 016801 (2002).
\bibitem{zhangnu014} Y. Zhang, Z. Jiang, J. P. Small, M. S. Purewal, Y. ÐW. Tan, M. Fazlollahi, J. D. Chudow, J. A. Jaszaczak, H. L. Stormer, and P. Kim, Phys. Rev. Lett. {\bf 96}, 136806 (2006).
\bibitem{lauaspectrat} F. Miao, S. Wijeratne, Y. Zhang, U. C. Coskun, W. Bao, and C. N. Lau, Science {\bf 317}, 1530 (2007).
\bibitem{JHChen08} J.-H. Chen, C. Jang, S. Adam, M. S. Fuhrer, E. D. Williams, and M. Ishigami, Nature Phys. {\bf 4}, 377 (2008).

\bibitem{Huard08} B. Huard, N. Stander, J. A. Sulpizio, and D. Goldhaber-Gordon, Phys. Rev. B. {\bf 78}, 121402(R) (2008).
\bibitem{blakecontact} P. Blake, R. Wang, S. V. Morozov, F. Schedin, L. A. Ponomarenko, A. A. Zhukov, R. R. Nair, I. V. Grigorieva, K. S. Novoselov, and A. K. Geim, Solid State Commun. {\bf 149}, 1068 (2009).

\bibitem{Bolotin08} K. I. Bolotin, K. J. Sikes, Z. Jiang, G. Fundenberg, J. Hone, P. Kim, and H. L. Stormer, Solid State Commun. {\bf 146}, 351 (2008).

\bibitem{Tan07} Y.-W. Tan, Y. Zhang, K. Bolotin, Y. Zhao, S. Adam, E. H. Hwang, S. Das Sarma, H. L. Stormer, and P. Kim, Phys. Rev. Lett. {\bf 99}, 246803 (2007).

\bibitem{Adam69} S. F. Adam, \textit{Microwave Theory and Applications} (Prentice-Hall, 1969).
\bibitem{sonnet} SONNET, Sonnet Software, Liverpool, NY, 2005.

\bibitem{Sutono01} A. Sutono, G. Cafaro, J. Laskar and M. Tentzeris, IEEE Trans. On Adv. Packaging {\bf 24}, 595 (2001).

\bibitem{Abanin08} D. A. Abanin and L. S. Levitov, Phys. Rev. B. {\bf 78}, 035416 (2008).

\bibitem{Hwang08} E. H. Hwang and S. Das Sarma, Phys. Rev. B. {\bf 77}, 195412 (2008).
\bibitem{Hong09}X. Hong, K. Zou, and J. Zhu,  Phys. Rev. B, {\bf 80} 241415 (2009).

\bibitem{Jiang07} Z. Jiang, E. A. Henriksen, L. C. Tung, Y. J. Wang, M. E. Schwartz, M. Y. Han, P. Kim, and H. L. Stormer, Phys. Rev. Lett. {\bf 98}, 197403 (2007).

\bibitem{McClure56} J. W. McClure, Phys. Rev. {\bf 104}, 666 (1956).
\bibitem{Haldane88} F. D. M. Haldane, Phys. Rev. Lett. {\bf 61}, 2015 (1988).
\bibitem{Iyengar07} A. Iyengar, J. Wang, H. A. Fertig, and L. Brey, Phys. Rev. B {\bf 75}, 125430 (2007).

\bibitem{Giesbers09} A. J. M. Giesbers, U. Zeitler, L. A. Ponomarenko, R. Yang, K. S. Novoselov, A. K. Geim, and J. C. Maan, Phys. Rev. B, {\bf 80} 241411 (2009).

\bibitem{sondhicarinirmp} S. L.  Sondhi, S. M. Girvin, J. P. Carini, and D. Shahar, Rev. Mod. Phys. {\bf 69} 315 (1997).

\end{thebibliography}
\end{document}